# Wave breaking on the surface of a dielectric liquid in a horizontal electric field

**Evgeny A. Kochurin and Olga V. Zubareva**
Institute of Electrophysics, Ural Branch of Russian Academy of Sciences
620016, 106 Amundsen str., Yekaterinburg, Russia

**Nikolay M. Zubarev**
Institute of Electrophysics, Ural Branch of Russian Academy of Sciences
620016, 106 Amundsen str., Yekaterinburg, Russia
P.N. Lebedev Physical Institute, Russian Academy of Sciences
119991, 53 Leninskij prospect, Moscow, Russia

**ABSTRACT**
The weakly nonlinear dynamics of the free surface of a dielectric liquid in an electric field directed tangentially to the unperturbed boundary is investigated numerically. Within the framework of the strong field model, when the effects of capillarity and gravity are not taken into account, it is shown that nonlinear surface waves have a tendency to break. In result of the collapse of the surface waves, the curvature of the boundary and the gradient of the local electric field undergo infinite discontinuity on the surface of the liquid. The angles of the boundary inclination remain small. The characteristic collapse time of a surface wave traveling in a given direction is calculated versus the dielectric constant of the liquid. It is shown that the time of the singularity formation increases infinitely at small and high values of the dielectric constant. The first case corresponds to the transition of the system to the neutral stability regime (the jump in electrostatic pressure at the boundary turns into zero). At high values of the dielectric constant of the liquid, the collapse time also increases. This effect is associated with the realization of a special regime of fluid motion, in which the propagation of nonlinear surface waves of an arbitrary configuration occurs without distortions. For the liquids with relative permittivity close to five, the wave breaking time reaches a minimum, i.e., the collapse of the surface waves for such liquids occurs most intensively.

## 1 INTRODUCTION

**THE** present work is devoted to the study of the nonlinear dynamics of the free surface of a non-conducting liquid in an external electric field directed tangentially to the unperturbed boundary. In such geometry, the external field stabilizes the surface [1]. The electric field directed perpendicularly to the unperturbed surface leads to the development of electro-hydrodynamic instability of the boundary [2]. Currently, the nonlinear dynamics of the boundaries of non-conducting liquids in external fields is actively studied both experimentally [3-5] and theoretically [6, 7]. Thus, the external electromagnetic fields can be used for controlling and stabilizing the dynamics of the fluid boundaries [8, 9].

In many analytical models of the nonlinear evolution of liquids with free and contact boundaries, singularities can arise in a finite time. From a physical point of view, the appearance of singularities corresponds to the formation of discontinuities in various quantities at the fluid boundary. Currently, it is customary to divide the forming singularities into two types: weak and strong ones. In the first case, the angles of the boundary inclination remain small, and, for example, the curvature of the surface can turn to infinity. Weak singularities can arise during the development of Kelvin-Helmholtz instability [10, 11], Tonks-Frenkel instability [12], and inertial motion of a free boundary of a liquid [13] (gravitational and capillary effects are negligibly small compared to a destabilizing factor). For strong singularities, the steepness of the boundary (the tangent of the angle of inclination to the surface) becomes discontinuous. Cusps, cones, and angles are the examples of such singularities. Strong discontinuities can form on the fluid boundaries in the gravitational force [14] and under the action of a strong vertical electric field [15, 16].

The liquid boundary in external fields can demonstrate not only singular, but also regular behavior. In this case, stable nonlinear waves that retain their shape (the so-called progressive waves) can propagate along the fluid surface, see, for example, [17, 18]. The theoretical description of the progressive waves has usually significant restrictions on the shape of the nonlinear perturbations under consideration (see book [19]). It was discovered in [6, 20, 21] that nonlinear waves of an arbitrary shape can propagate without distortions along the surface of a non-conducting liquid with high dielectric constant in a strong tangential electric field. The fact that the geometry of such waves is not restricted distinguishes them from other types of progressive waves and makes them a very interesting object of study [22].

Exact solutions [6] were obtained for the special case of the liquid with high dielectric constant. Despite the fact that a sufficiently wide class of substances (distilled water and various alcohols) can be attributed to the liquids with high permittivities, this condition severely restricts the generality of the problem. The aim of this work is to study the nonlinear dynamics of the boundaries of liquids in a strong horizontal electric field without restrictions on their permittivities. In the framework of numerical methods, it will be shown that in the case of a finite dielectric constant, weakly nonlinear waves at the fluid boundary tend to break. The collapse of a surface wave corresponds to the formation of a weak singularity for which the curvature of the boundary and the gradient of the local electric field become infinite. The angles of inclination of the boundary remain small. In the present work, the wave breaking process is studied for liquids with different dielectric constants. In particular, it is analytically and numerically shown that the collapse of surface waves occurs most rapidly for the liquids with a relative permittivity close to five. Thus, in this work, a new physically distinguished case is discovered in which the collapse of surface perturbations in a strong horizontal electric field occurs most intensively.

## 2 LINEAR ANALYSIS OF THE PROBLEM

Let us consider an evolution of an ideal incompressible non-conducting fluid with a free surface in an external electric field directed tangentially to the unperturbed boundary. We assume that the liquid is deep, i.e., the wavelength of surface perturbations is much less than the depth of the liquid. Since the problem under consideration is anisotropic, we restrict our consideration only by plane waves traveling in the direction of the external horizontal field. Let the vector of an electric field strength be directed along the $x$ axis (correspondingly, the $y$ axis of the Cartesian coordinate system is perpendicular to it and to unperturbed boundary) and has the absolute value $E$. The surface profile is described by the function $\eta(x, t)$; for the unperturbed state $\eta = 0$.

The dynamics of linear waves propagating along the surface of a non-conducting liquid under the action of the tangential electric field is described by the following dispersion relation [1, 2]:

$$\omega^2 = gk + \frac{\varepsilon_0 \gamma(\varepsilon)}{\rho} E^2 k^2 + \frac{\sigma}{\rho} k^3, \qquad (1)$$

where $\omega$ is the wave frequency, $k$ is the wavenumber, $g$ is the gravity acceleration, $\varepsilon_0$ is the dielectric constant of vacuum, $\gamma(\varepsilon) = (\varepsilon-1)^2/(\varepsilon+1)$ is the auxiliary coefficient, $\varepsilon$ and $\rho$ are the dielectric constant and the mass density of the fluid, respectively, and $\sigma$ is the surface tension coefficient.

In a situation when the external field is absent, the dispersion relation (1) describes the dynamics of linear capillary-gravity waves. The phase speed of such waves has a minimum at the wavenumber $k_0 = (g\rho/\sigma)^{1/2}$. The minimum value of the velocity is $V_0 = (4\sigma g/\rho)^{1/4}$. It means in the absence of the external field, the surfaces waves do not propagate with the velocity less than $V_0$. It is possible to introduce the characteristic value of the electric field strength as

$$E_0^2 = (\sigma g \rho)^{1/2}/\varepsilon_0 \gamma. \qquad (2)$$

In such a field, all terms on the right-hand side of Equation (1) make the same contribution to the phase velocity at $k = k_0$. The characteristic field $E_0$ decreases with increase in dielectric constant. The minimal value of the field is reached for the liquids with high $\varepsilon$. For water ($\varepsilon \approx 80$), it is estimated as 2 kV/cm. For a liquid with low $\varepsilon$, like diesel oil (used in the experimental research [3, 4]) with $\varepsilon \approx 2.2$, the field should be near 17 kV/cm.

The characteristic scales of the length and time are

$$\lambda_0 = 2\pi \left(\frac{\sigma}{g\rho}\right)^{1/2}, \quad t_0 = 2\pi \left(\frac{\sigma}{g^3\rho}\right)^{1/4}. \qquad (3)$$

For most liquids, these values can be estimated in order of magnitude as $\lambda_0 \approx 1$ cm, and $t_0 \approx 0.1$ s.

In a situation when the external field is strong enough, the term responsible for the influence of electrostatic force dominates over gravity and capillary ones. For the strong field case, Zubarev [6] has found exact analytical solutions of the full electro-hydrodynamic equations, which are applicable for the liquids with high dielectric constant, $\varepsilon \gg 1$. According to these solutions, the strongly nonlinear waves of any shape can propagate without dispersion in the direction, or against the direction of the tangential electric field. Only interaction between oppositely-traveling waves leads to the formation of singular points on the surface of the liquid [23, 24].

## 3 EQUATIONS OF MOTION

In this section, we write the equations system of electro-hydrodynamics. We introduce the velocity potential $\phi(x, y)$, defining the velocity of the fluid $V(x,y) = \nabla \phi$, where $\nabla = \{\partial/\partial_x, \partial/\partial_y\}$ is the nabla-operator. We also introduce the electric field potentials inside ($\varphi_1$) and outside ($\varphi_2$) the liquid. The velocity and electric field potentials obey the Laplace equations:

$$\nabla^2 \phi = 0, \qquad \nabla^2 \varphi_{1,2} = 0.$$

On the free surface of the liquid, the dynamic boundary condition is satisfied:

$$\phi_t + (\nabla\phi)^2/2 = -(P_0 - P_E)/\rho - g\eta + \sigma K/\rho, \quad y = \eta, \quad (4)$$

where $P_E$ is the pressure exerted by an external electric field, $P_0$ is a constant, $K = \eta_{xx}/(1+\eta_x^2)^{3/2}$ is the boundary curvature (we introduce the notation for the partial derivative as, $f_x = \partial f / \partial x$). The quantities $P_E$ and $P_0$ have the following form:

$$P_E = \varepsilon_0(\varepsilon-1)(\nabla\varphi_1 \cdot \nabla\varphi_2)/2, \qquad P_0 = \varepsilon_0(\varepsilon-1)E^2/2.$$

The condition that the fluid does not flow through its boundary (kinematic boundary condition) should be satisfied together with the dynamic boundary condition (4)

$$\eta_t = \phi_y - \eta_x \phi_x, \qquad y = \eta. \quad (5)$$

As the liquid is ideal dielectric, the electric field potentials satisfy the boundary conditions at $y = \eta(x, t)$: $\varphi_1 = \varphi_2$ and $\varepsilon \partial_n \varphi_1 = \partial_n \varphi_2$, where $\partial_n$ denotes the derivative along the normal to the free surface. At infinite distance from the surface, $y \to \mp\infty$, the electric field potentials describe the stationary homogeneous electric field distribution, $\varphi_{1,2} \to -Ex$. At $y \to -\infty$, the fluid is at rest, i.e., $\phi \to 0$.

These equations are a closed equations system describing the fully nonlinear evolution of a dielectric liquid with free surface taking into accounts the effects of gravitational, capillary, and electrostatic forces. In the linear approximation, when the amplitude of the surface waves is negligibly small, the system turns into the dispersion equation (1).

## 4 WEAKLY NONLINEAR MODEL

For the convenience of further analysis, let introduce the dimensionless variables:

$$\eta \to \eta \cdot \lambda_0, \quad x \to x \cdot \lambda_0, \quad t \to t \cdot t_0, \quad \phi \to \phi \cdot \lambda_0^2/t_0,$$

where $\lambda_0$, $t_0$ are the characteristic values of the length and time defined by (3). It is convenient to introduce the dimensionless parameter $\beta$ defining the electric field strength as follows $\beta^2 = \varepsilon_0 E^2 (\sigma g \rho)^{-1/2} = E^2/\gamma E_0^2$, where $E_0$ is determined by (2), i.e., if $E = E_0$ then $\beta^2 = 1/\gamma$. Here, and further, we assume that the electric field is strong enough, i.e., $\gamma \beta^2 \gg 1$. In such a situation, we can neglect the influence of capillary and gravity effects. Then, in the dimensionless units, the dispersion relation (1) takes the simple form,

$$\omega^2 = \gamma \beta^2 k^2, \quad (6)$$

in the following range of wavenumbers

$$1/\gamma \beta^2 \ll k \ll \gamma \beta^2.$$

The relation (6) describes the dispersionless propagation of the linear surface waves with the velocity $\gamma^{1/2}\beta$.

In the present work, the weakly nonlinear dynamics of the free surface of a liquid dielectric is considered, i.e., the angles of the boundary inclination are small, $|\eta_x| \sim \alpha \ll 1$ ($\alpha$ is the smallness parameter). As was shown in [6], the electro-hydrodynamics equations presented in the previous section can be reduced to the simpler equations system describing directly the nonlinear dynamics of the surface. Using the expansion in $\alpha$, we can write the resulting equations system including the quadratically nonlinear terms:

$$\psi_t = -\gamma \beta^2 \hat{k}\eta + \frac{1}{2}\left[\gamma \beta^2 A_E[(\hat{k}\eta)^2 - (\eta_x)^2] + (\hat{k}\psi)^2 - (\psi_x)^2\right]$$
$$+ \gamma \beta^2 A_E \left[\hat{k}(\eta\hat{k}\eta) + (\eta\eta_x)_x\right], \quad (7)$$

$$\eta_t = \hat{k}\psi - \hat{k}(\eta\hat{k}\psi) - (\eta\psi_x)_x, \quad (8)$$

where $\psi(x,t) = \phi(x,\eta(x,t),t)$ is the value of the velocity potential at the boundary of the liquid, $\hat{k}$ is the integral operator having the form $\hat{k} f_k = |k| f_k$ in the Fourier representation. Here $A_E$ is the electric analogue of the Atwood number defined as

$$A_E = (\varepsilon-1)/(\varepsilon+1).$$

Equations (7) and (8) correspond to the weakly nonlinear form of Equations (4) and (5), respectively. The term containing the electric field potentials has been excluded from Equation (4) using the relation [6]:

$$\varphi_1|_{y=\eta} = \varphi_2|_{y=\eta} = A_E \beta \left( \hat{H}\eta - A_E \left[ \eta \eta_x + \hat{H}(\eta \hat{H}\eta_x) \right] \right), \quad (9)$$

where $\hat{H}$ is Hilbert transform defined in Fourier-space as, $\hat{H}f = i\,\mathrm{sign}(k) f_k$. It is related with $\hat{k}$ operator: $\hat{k} = -\hat{H}\partial_x$.

The main goal of this work is numerical study of the system of the nonlinear integro-differential equations (7) and (8). It will be shown that in the case of a finite dielectric constant, Equations (7) and (8) describe the wave breaking process. In result of the collapse of surface waves, the discontinuities in the surface curvature and in the gradient of electric field form at the surface of the liquid. We present the results of numerical simulation in detail in the next section.

## 5 SIMULATION RESULTS

The numerical scheme for integrating Equations (7) and (8) is based on the pseudo-spectral methods; it means that the functions $\psi$ and $\eta$ are approximated by the finite Fourier series. Consequently, the boundary conditions for Equations (7) and (8) are periodic in space with the period $2\pi$. The total amount of the Fourier harmonics used in calculation presented was $N = 2^{15}$. For the numerical integration in time, we use the explicit fourth-order Runge-Kutta method with the step $dt = 10^{-6}$. The initial conditions for (7) and (8) are taken in the form:

$$\eta(x,0) = a\cos(x), \qquad \psi(x,0) = -a\gamma^{1/2}\beta \sin(x). \quad (10)$$

In the linear approximation, these expressions describe the propagation of a periodic surface wave with the amplitude $a$ in the direction opposite to the $x$ axis.

We present the results of simulations for the parameters: $a = 0.05$, $\beta = 10$, $\varepsilon = 2.2$. Such the value of dielectric constant is selected, because it corresponds to diesel oil used in the experimental researches [3, 4], in terms of $\gamma$, it corresponds to $\gamma = 0.45$. Figure 1 shows the shape of the surface at the initial time and in the end of the calculation interval $t_c \approx 4.44$. The computations were stopped when the relative error in calculation of the system energy reached the value of $10^{-9}$.

From Figure 1, it can be seen that a weakly nonlinear wave in the external horizontal electric field tends to break. The main question is whether the conditions for the applicability of the weakly nonlinear approximation are satisfied for this process. Figure 2 shows the steepness of the fluid boundary (the first spatial derivative) at the initial time and at the end of the calculation interval. It can be seen that a region ($x \approx 0.22$) with a sharp drop in the steepness of the boundary has formed on the liquid surface. At the same time, the angles of the boundary inclination remain relatively small; the maximum value of $\eta_x$ does not exceed 0.15. Such a behavior of the system may indicate the formation of an infinite discontinuity in the curvature of the surface.

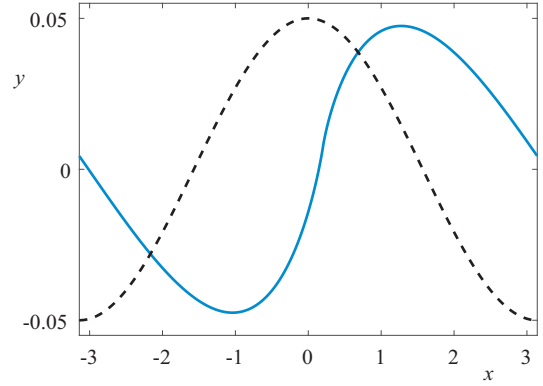

**Figure 1.** The surface of the liquid is shown at the initial moment (black dashed line) and at the end of the computational interval (blue solid line) $t_c \approx 4.44$.

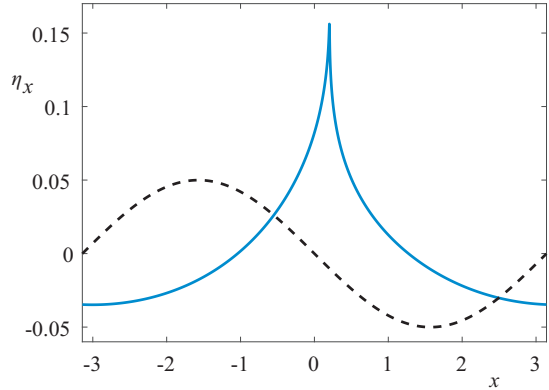

**Figure 2.** The steepness of the liquid surface is shown at the initial moment (black dashed line) and at the end of the computational interval (blue solid line) $t_c \approx 4.44$.

Figure 3 shows the curvature of the boundary at the initial moment and in the end of the calculation interval. Indeed, it can be seen that at the point $x \approx 0.22$, the curvature of the boundary increases significantly undergoing a discontinuity. In general, the observed behavior of the system is very similar to the formation of weak root singularities [10-13]. The curvature of the boundary near the root singularity is described by the power-law function [10]:

$$K \approx \eta_{xx} \sim |x - x_0|^{-1/2}, \quad (11)$$

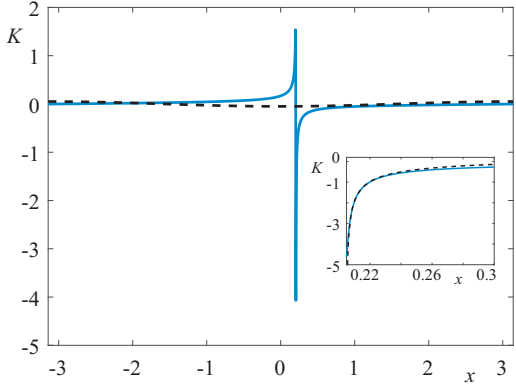

**Figure 3.** The surface curvature is shown at the initial moment (black dashed line) and at the end of the computational interval (blue solid line) $t_c \approx 4.44$. The inset to the figure shows the curvature (blue solid line) at $t_c$ and the power-law dependence (11) (black dashed line).

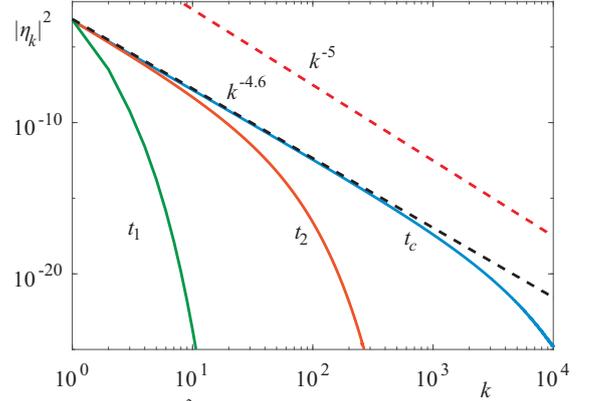

**Figure 4.** Spectra $|\eta_k|^2$ are shown at successive times $t_1 = 0.4$, $t_2 = 4.00$, and $t_c = 4.44$. The black and red dashed lines correspond to the power dependences $k^{-4.6}$ and $k^{-5}$, respectively.

where $x_0$ is the coordinate of the singularity, in the current case $x_0 \approx 0.22$. The inset to the Figure 3 shows the curvature of the surface on the interval $0.22 \leq x \leq 03$. The black dashed curve in the inset corresponds to the power-law dependence (11). It can be seen that the power-law dependence (11) very well describes the calculated behavior of the surface curvature near the point $x_0$.

It should be noted that the appearance of a discontinuity in the curvature of the boundary corresponds to the formation of a singularity in the gradient of the electric field strength on the surface. According to (9), the local electric field is proportional to $\eta_x$ and, consequently, the gradient of the field on the surface of fluid is determined by the value of the curvature of the boundary $\eta_{xx}$. Thus, the formulas (9) and (11) describe the formation of infinite discontinuities in the boundary curvature and in the gradient of the local electric field on the surface of fluid.

The spectra of the surface elevation $|\eta_k|^2$ also indicate to the singular behavior of the system; they are presented at successive instants of time in Figure 4. The Fourier spectrum of a smooth function which has no geometrical singularities has an exponentially decaying character. Figure 4 shows that, in the end of the calculation interval, the surface spectrum has not exponential but the power-law dependence: $|\eta_k|^2 \sim k^{-4.6}$, which evidences on the singular behavior of the system. It should be noted the Fourier spectrum of the surface for the curvature singularity (11) decays as $k^{-5}$ [10], which is close to the observed spectrum.

The next question is how the parameters of the wave breaking depend on the value of the dielectric constant of the liquid. Figure 5 shows the time $t_c$, for which the initial perturbation (10) collapses to the weak root singularity. The initial wave amplitude $a$ in (10) was chosen to be 0.1, the dimensionless field was the same as in the previous simulation, $\beta = 10$. The calculations were carried out for different values of the dielectric constant. For clarity of the results obtained, the collapse time $t_c$ in Figure 5 is shown versus the electric Atwood number $A_E$, the values of which are restricted by the interval: $0 < A_E < 1$. In terms of the dielectric constant, the case of $A_E = 0$ corresponds to $\varepsilon = 1$; it describes the neutral stability regime, i.e., the electrostatic pressure jump is absent on the surface. The opposite case of the high dielectric constant ($A_E = 1$) corresponds to the Zubarev's exact solution [6], describing the propagation of the nonlinear wave without distortions. From Figure 5, we can see that, for the both limiting cases $A_E \to 0$ and $A_E \to 1$, the collapse time increases infinitely. The minimum value of $t_c$ is reached near

$$A_E \approx 0.7, \qquad \varepsilon \approx 5.67. \qquad (12)$$

A type of the observed singularities did not depend on the values of the electric field strength and dielectric constant of the liquid. In the wave breaking process, infinite discontinuities in the boundary curvature and in the gradient of the electric field on the surface were formed. The angles of the boundary inclination remain small. Thus, the applicability criteria of the model were not broken during the formation of the observed weak singularities.

The calculated dependence $t_c(A_E)$ has the following two asymptotes for the small and high values of the dielectric constant:

$$t_c \sim A_E^{-1}, \qquad A_E \to 0, \qquad (13)$$

$$t_c \sim (1-A_E)^{-0.3}, \qquad A_E \to 1. \qquad (14)$$

The results presented in Figure 5 were obtained for the particular case of the initial surface perturbation in the form (10). We now analytically demonstrate that the characteristic time of the wave collapse is universal and does not depend on the initial waveform. Let us define a universal nonlinear time $\tau_c$ describing the deformation of a traveling surface wave. We rewrite Equations (7) and (8) in the form describing the evolution of a wave traveling in a given direction. Let introduce new unknown function $f(x, t)$

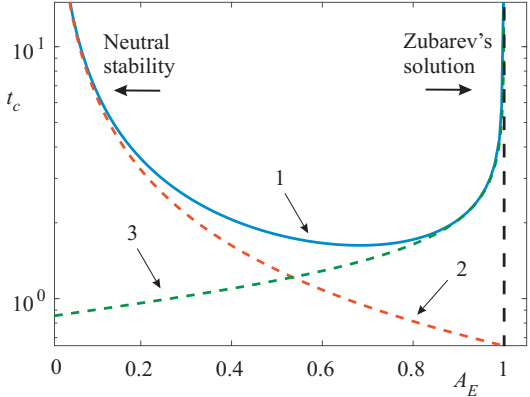

**Figure 5.** The time of the discontinuity formation is shown versus the electric Atwood number $A_E$. The blue solid line "1" corresponds to the calculation results, the red "2" and green "3" dashed lines are the asymptotes $0.65 \cdot A_E^{-1}$ and $1.05 \cdot (1 - A_E)^{-0.3}$, respectively.

(which has a sense of the wave traveling in the opposite direction to *x* axis):

$$f(x,t) = \frac{1}{2}\left[\eta - (\gamma^{1/2}\beta)^{-1}\hat{H}\psi\right],$$

and new independent variables $u = x + (\gamma^{1/2}\beta)t$, and $\tau = \alpha t$. From the physical point of view, such replacement of variables corresponds to the transition to a moving frame of reference in which the wave $f(u, \tau)$ does not change its shape in the linear approximation. The nonlinear evolution is described in slow time $\tau$ by the single nonlinear equation; see for more details [6]:

$$f_\tau = \frac{\gamma^{1/2}(1-A_E)\beta}{2}\left(\hat{H}(ff_u)_u - f_u\hat{H}f_u - (f\hat{H}f_u)_u\right). \quad (15)$$

Equation (15) is obtained from the equations (7) and (8) under the assumption that the amplitude of the waves traveling in the direction of the *x*-axis is negligible small. The coefficient before the nonlinear term in right-hand side of Equation (15) has a sense of the inverse characteristic time in the problem under study. It is possible to rewrite this time in terms of the electric Atwood number:

$$\tau_c = \frac{2}{\gamma^{1/2}(1-A_E)\beta} \sim \frac{1}{A_E(1-A_E)^{1/2}}. \quad (16)$$

The quantity $\tau_c$ in the formula (16) demonstrates a qualitatively similar behavior as the calculated time $t_c$ plotted in Figure 5. The time $\tau_c$ infinitely increases for the small and high values of $A_E$. As well as the calculated time $t_c$, the quantity $\tau_c$ has a minimum. It reaches at the following value of the electric Atwood number:

$$A_E = 2/3, \qquad \varepsilon = 5, \qquad (17)$$

which is in very good agreement with the numerical data (12). It should be noted that the expression (17) is exact. It means that this value of dielectric constant, i.e., $\varepsilon = 5$, is physically distinguished. For $\varepsilon = 5$, the surface waves break most intensively.

For the small values of $A_E$, the formula (16) gives exactly the asymptote (13). At the same time, for the high dielectric constant, the relation (16) has the asymptote $\tau_c \sim (1-A_E)^{-0.5}$, which is slightly different from the calculated asymptote (14). The difference in the asymptotic exponents may be due to the finite accuracy of the numerical methods used and to the influence of high order nonlinear effects not taken into account in Equation (15).

## 6 CONCLUSION

The nonlinear dynamics of the free surface of a liquid dielectric in a strong horizontal electric field is numerically studied in the framework of the weakly nonlinear model. Computational data show that nonlinear surface waves break: in result of the wave collapse, discontinuities in the boundary curvature and in the gradient of the local electric field form on the surface of the liquid. The angles of the boundary inclination remain small. Thus, the applicability criteria of the model were not broken during the computations carried out.

The observed behavior of the fluid boundary is similar to the formation of weak root singularities arising in the different physical situations [10-13]. The spectrum of the surface elevation in the moment of the wave collapse has a power-law dependence, which indicates to the singular behavior of the system. The exponent of the surface spectrum tends to the value −4.6, which is close to the exponent of the weak root singularity (minus five) [10].

The dependence of the singularity formation time on the dielectric constant of the liquid was calculated. The obtained dependence demonstrates two different regimes of an asymptotic behavior at low and high values of the dielectric constant. With a decrease in the dielectric constant, the time of the singularity formation increases infinitely. This situation corresponds to the transition to the regime of the neutral stability of the liquid. At high values of the permittivity, the wave collapse time also increases. This effect is associated with the realization of the regime of the distortionless propagation of nonlinear waves, which was first discovered in [6]. It is analytically and numerically shown that the time of the wave collapse has a minimum value for the liquids with the relative dielectric constant equal to 5 (for a fixed electric field strength).

In conclusion, we note that the process of the discontinuity formation triggers the transitions of the system energy to the small scales. In the framework of a model accounting the effects of surface tension, the capillary waves will be generated in the regions of high boundary curvature. Small-scale capillary waves will be emitted until their wavelength reaches the values at which the influence of viscosity will dominate. Thus, the wave breaking mechanism in the strong field limit investigated in this work can lead to the development of the wave turbulence on the liquid surface in the finite field case [5, 24, 25].

## ACKNOWLEDGMENT

The reported study was funded by RFBR, project number 20-38-70022.

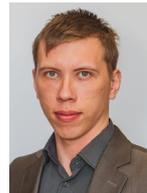

**Evgeny A. Kochurin** was born in 1988. He received M.Sc. degree in electrical physics from Ural Federal University, in 2011. Since 2010 he has been employed at the Institute of Electrophysics, Ural Branch of Russian Academy of Sciences, Yekaterinburg city. He received Candidate of Science degree (PhD) form the Institute, in June, 2015. His PhD thesis was devoted to theoretical investigation of nonlinear dynamics of free and contact boundaries of dielectric liquids under the action of strong electric field.

**Olga V. Zubareva** was born in 1971. In 1994, she graduated from the Physics department of Ural State University. Since 1994 she has been employed at the Institute of Electrophysics, Yekaterinburg, Russia. She received the Candidate of Science degree (PhD) from the Institute, in 2002. She works as a senior research scientist in the nonlinear dynamics laboratory. She is involved in theoretical studying of structures on the surface of fluids in electric and magnetic fields, conditions of their existence and stability.

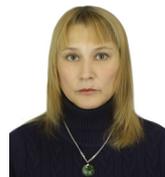

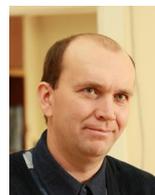

**Nikolay M. Zubarev** was born in 1971. He received the M.Sc. degree in applied mathematics and physics from Moscow Institute of Physics and Technology, Moscow, Russia, in 1994, the Cand.Sci. degree from the Institute of High Current Electronics, Russian Academy of Sciences, Tomsk, Russia, in 1997, and Dr.Sci. degree from the Institute of Electrophysics, Russian Academy of Sciences, Yekaterinburg, Russia, in 2003. He is currently with the Institute of Electrophysics, Russian Academy of Sciences, where he is involved in theoretical studying of nonlinear phenomena in liquids with free surface under the action of an electric field and electrical discharges in gas and vacuum. Prof. Zubarev is a Corresponding Member of the Russian Academy of Sciences since 2016. He was a recipient of the State Prize of the Russian Federation in Science in 2003.